\documentclass[3p]{elsarticle}

\usepackage{amssymb,euscript,latexsym,amsmath}
\usepackage{graphicx}
\usepackage[dvips]{epsfig}
\usepackage{color}
\usepackage{epstopdf}
\usepackage{setspace}
\usepackage{subfigure} 
\usepackage{lineno,hyperref}
\modulolinenumbers[5]

\journal{Journal of Biomechanics}

\DeclareGraphicsExtensions{.eps,.ps,.pdf,.tif}
\graphicspath{{NewFigures/}}

\begin{document}

\begin{frontmatter}

\title{Slow Limb Movements Require Precise Control of Muscle Stiffness}


\author[mymainaddress,mythirdaddress]{Sarine Babikian}

\author[mymainaddress,mysecondaryaddress]{Francisco J. Valero-Cuevas\fnref{footnote1}
}

\author[mythirdaddress]{Eva Kanso\corref{mycorrespondingauthor1}\fnref{footnote1}
}
\cortext[mycorrespondingauthor1]{Corresponding author}
\ead{kanso@usc.edu}

\address[mymainaddress]{Biomedical Engineering, University of Southern California, Los Angeles, California, USA}
\address[mysecondaryaddress]{Biokinesiology and Physical Therapy, University of Southern California, Los Angeles, California, USA}
\address[mythirdaddress]{Aerospace and Mechanical Engineering, University of Southern California, Los Angeles, California, USA}

\fntext[footnote1]{Valero-Cuevas and Kanso have equal leadership of this paper.}

\begin{abstract}

Slow and accurate finger and limb movements are essential to daily activities, but their neural control and governing mechanics are relatively unexplored. We consider neuromechanical systems where slow movements are produced by neural commands that modulate muscle stiffness. This formulation based on strain-energy equilibria is in agreement with prior work on neural control of muscle and limb impedance. Slow limb movements are \textit{driftless} in the sense that movement stops when neural commands stop. We demonstrate, in the context of two planar tendon-driven systems representing a finger and a leg, that the control of muscle stiffness suffices to produce stable and accurate limb postures and quasi-static (slow) transitions among them. We prove, however, that stable postures are achievable only when muscles are pre-tensioned, as is the case for natural muscle tone. Our results further indicate, in accordance with experimental findings, that slow movements are non-smooth. The non-smoothness arises because the precision with which individual muscle stiffnesses need to be controlled changes substantially throughout the limb's motion. These results underscore the fundamental roles of muscle tone and accurate neural control of muscle stiffness in producing stable limb postures and slow movements. 

\end{abstract}


\end{frontmatter}



\section*{Introduction}

Fine grasp and manipulation of objects such as threading a needle or writing require accurate finger and limb postures and slow movements. Inaccuracies in performing such tasks often arise in neuromuscular conditions such as spasticity, Parkinson's disease and stroke~\cite{Bettray2013, Krebs1999}. An understanding of these inaccuracies and their causes necessitates a deeper knowledge of the neuromechanical conditions required for the muscles to produce slow and accurate limb movement. The neural control of muscles is complex~\cite{Hogan1984}; however, it is believed that the neural input determines the force and stiffness of the muscle \cite{Ford1977, Gordon1966, Mason1978, Morgan1977, Rack1974, Cooke1980}. We therefore propose a computational framework for slow movements where the neural control of muscle activity is simulated by altering the muscle stiffness properties. 

Slow limb movements  belong to a class of mechanical systems, known as \textit{driftless}, where movement stops when actuation stops. Driftless systems arise in many applications, including  satellite dynamics~\cite{Teel1995, Boscain2002}, robotic vehicles~\cite{De-Luca1998}, and  biolocomotion~\cite{Shapere1989, Kanso2005, Kanso2009, Jing2013}. We apply these principles to examine the neuromechanical properties of \textit{quasi-static} limb movements. This viewpoint is reminiscent of but not identical to the \textit{equilibrium point hypothesis} (EPH), which states that limb movements are generated by a sequence of equilibrium points along a desired trajectory~\cite{Asatryan1965, Feldman2009, Cooke1980}. EPH considers that the magnitude of the ``control" force exerted on the limb, at any time, depends on the difference between the ``actual" limb dynamics and the desired equilibrium point, and derives a second-order differential equation for the ``error" between the actual and desired dynamics~\cite{Shadmehr1998, Asatryan1965, Feldman1966}. Here, we are not concerned with the relaxation dynamics. We posit that this relaxation time scale is much faster than the time scale associated with slow limb movements, and the limb reaches a state of equilibrium instantaneously in response to changes in muscle stiffness. We seek to determine whether equilibrium postures of multi-joint limbs and quasi-static transitions among them can be achieved by modulating the muscles' stiffness values.

We demonstrate that the control of muscle stiffness suffices to produce stable limb postures and quasi-static transitions among them. This is in agreement with results from impedance control~\cite{Hogan1985, Mussa-Ivaldi1991}. However, we go beyond these results to prove that stable postures are achievable only when muscles are pre-tensioned as in the case of natural muscle tone. 
Muscle tone is defined clinically as a state of active muscle contraction at rest~\cite{Sanger2003}, and depends on the physiological or pathological state of the limb. Healthy muscles at rest are always producing a baseline level of force at the tendon~\cite{Williamson2006}. This baseline force is equivalent to the muscle ``threshold" value of EPH~\cite{Feldman2009}. Unlike threshold control, we do not alter the pre-tensioning length directly.  Rather, we define muscle tone as a baseline non-zero tension in the tendon, and view muscle stiffness as a neurally controlled parameter.  

Controlling muscle stiffness to produce equilibrium limb postures does not admit a unique solution due to the over-actuated nature of the tendon-driven systems, thus, the same posture can be achieved at multiple combinations  of stiffness parameter values. Each combination has a distinct strain energy cost. If a relatively-low energy level is inaccessible, say due to a pathological condition, the same posture may be realized at a higher energy cost.  Therefore, we formulate the problem of muscle stiffness control as an optimization problem~\cite{Ivaldi1988, Mussa-Ivaldi1991} that  minimizes the strain energy of the limb. Our optimal stiffness values are in agreement with the reciprocal-inhibition phenomenon: relaxation of the antagonist muscle during activity of the agonist \cite{Sherrington1913}. Further, we find that, for constant joint moment arms and muscle pre-tensioning values, optimal stiffness values depend on the change in the joint angle but not on the reference posture.  This symmetry with respect to reference postures is broken in the case of the anatomically more realistic, posture-dependent moment arms.

Finally, we investigate the smoothness of quasi-static transitions among limb postures given physiological limits on the neural control of muscle stiffness. 
Discontinuities, or non-smoothness, in slow limb movements have been reported in experimental studies~\cite{Vallbo1993, Wessberg1995, Wessberg1996}, 
and are thought to be due to non-trivial coupling among multiple neural and peripheral sources \cite{Evans2003, Gross2002, Williams2009, Williams2010}. 
The intrinsic stiffness of individual muscles is linked to muscle activation and force~\cite{Cui2008, Hu2011, Burdet2013}. Therefore, the precision with which muscle stiffness can be controlled is in the same general range as the precision with which muscle force can be controlled. Prior work~\cite{Keenan2007, Moritz2005, Slifkin2000, Harris1998} shows that, at low force productions, muscle force can be regulated with a precision in the range of 2.5 to 10\%. 
Under such limitations, we observe discontinuities in the limb's quasi-static trajectories, indicating unreachable postures because of insufficient precision in the muscle stiffness parameters. Our results indicate that these discontinuities are spatially localized in the limb workspace.

\section*{Methods}

\begin{figure}[!t]
\begin{center}
       \includegraphics[width=1\textwidth]{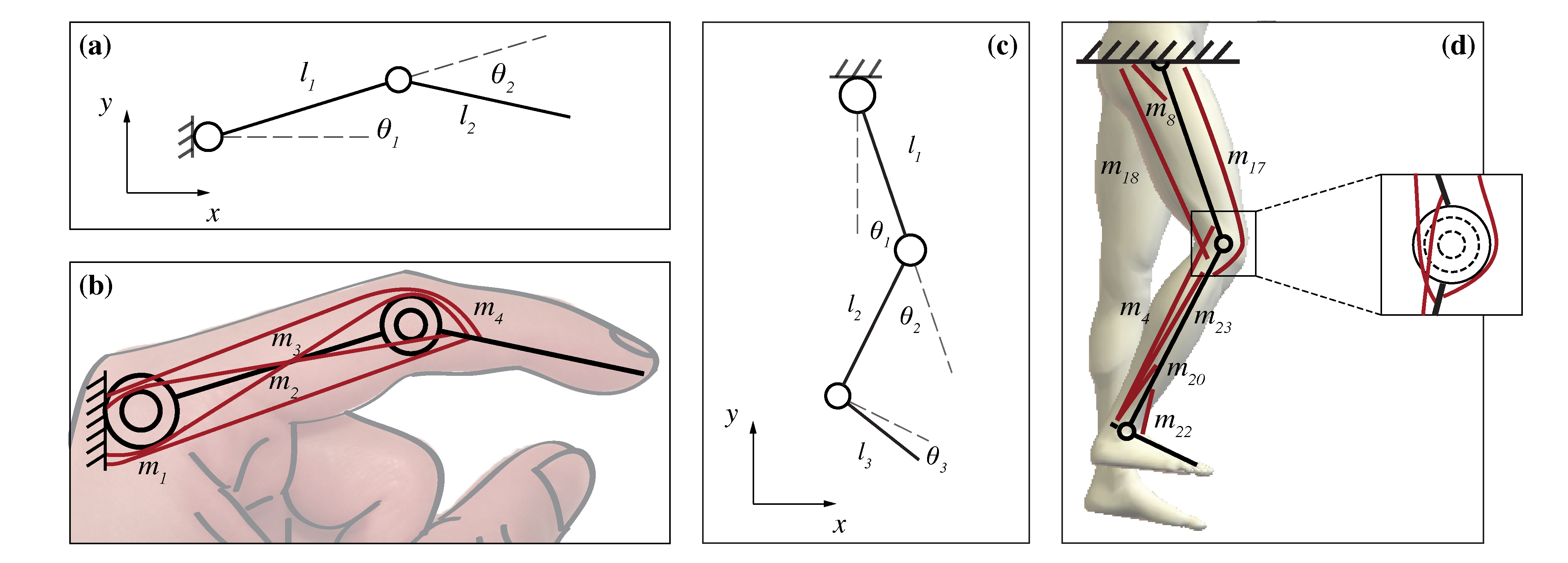}
 \end{center}
 \caption{Finger (a,b) and leg (c,d) models: kinematics (a,c) and muscle routing (b,d) overlaid on top of the physiological system.} 
  \label{fig:models}
\end{figure}

We examine the slow  movements of tendon-driven limbs in the context of two planar model systems: an idealized finger model and a more anatomically-realistic leg model. For notational convenience, we consider the general case of a planar limb of $n$ joints and let $l_1, \ldots, l_n$ be the lengths of the individual limb segments. The limb posture is defined by the column vector of joint angles $\boldsymbol{\theta}=(\theta_1, \ldots , \theta_n)^T$ where the superscript $()^T$ denotes the transpose operator. The joints are driven by $m$ muscles  ($m>n$) of variable stiffness parameters $k_i$, $i=1,\ldots,m$, whose tendon paths produce an $n\times m$ moment arm matrix $\mathbf{R}(\boldsymbol{\theta})$~\cite{Valero-Cuevas2009}, 
\begin{equation}\label{eq:Rmtx}
\mathbf{R}= 
\begin{pmatrix} 
r_{11} & r_{12} &  \cdots & r_{1m}\\  
\vdots & \vdots & \ddots & \vdots \\
r_{n1}&r_{n2}& \cdots&r_{nm}\\ 
\end{pmatrix}.
\end{equation}
Each entry $r_{ij}$ (here, $i=1,\dotsc , n$, $j=1,\dotsc , m$) denotes the ``rotation" the $j^{th}$ tendon produces at the $i^{th}$ joint. It is positive when pulling the $j^{th}$ tendon induces a positive  rotation (i.e., counterclockwise rotation per the right-hand rule). 
Depictions of the finger ($n=2$) and leg ($n=3$) models and their tendon paths are shown in Figure~\ref{fig:models}. Note that the extensor mechanism of the fingers is rather complex~\cite{Valero-Cuevas2007}, and Figure~\ref{fig:models}(b) corresponds to an idealized case of multi-articular muscles whose moment arms change signs. The tendon paths of the leg model in Figure~\ref{fig:models}(d) are more anatomically-realistic~\cite{Kuo1993,Kutch2012,Arnold2010}. For clarity, Figure~\ref{fig:models}(d) shows only seven of the fourteen muscles/muscle groups, amounting to a total of twenty-three individual muscles that we use in this leg model. 
In both the finger and leg models, we examine two types of joints: simple hinge joints where the moment arm matrix $\mathbf{R}$ is constant for all values of $\boldsymbol{\theta}$, and non-circular joints where the moment arm matrix is posture-dependent. 
In the finger model, the constant moment arm matrix is chosen to be

$\mathbf{R}= 
\left( \begin{array}{cccc} 
- 0.8 & - 0.8 & 0.7 & 0.7\\  
- 0.5 & 0.2 & -0.5 & 0.2 
\end{array} \right),$

in length units of cm. These values lie in the range of average moment arm values for the index finger reported in~\cite{An1983,Valero-Cuevas1998}. For the posture-dependent moment arms, we define joint ellipses with semi-major and semi-minor axes in the range of measured index moment arms found in~\cite{An1983}. In the leg model, we utilize the posture-dependent moment arm values of the model developed by~\cite{Arnold2010}.

A change in limb posture corresponds to a rotation $\Delta\boldsymbol{\theta}= \boldsymbol{\theta}_1- \boldsymbol{\theta}_0$ of the joints from a reference limb posture $\boldsymbol{\theta}_0$ to a new limb posture $\boldsymbol{\theta}_1$. A rotation $\Delta\boldsymbol{\theta}$ fully determines the excursions $\Delta\mathbf{s} = (\Delta s_1, \ldots,  \Delta s_m)^T$ of all muscles~\cite{An1983, Valero-Cuevas2009a},
\begin{equation}\label{eq:excursions}
\Delta\mathbf{s}=-\int_{\boldsymbol{\theta}_0}^{\boldsymbol{\theta}_1}
\left(\mathbf{R}(\boldsymbol{\theta})\right)^T \rm{d} \boldsymbol{\theta} = -\mathbf{R}^T \Delta\boldsymbol{\theta}~ , 
\end{equation}
where the negative sign indicates that a positive rotation of the joint will shorten the muscles that induce it; and vice versa. 
For constant moment arms, the second equality in~\eqref{eq:excursions} is evident. In the case of posture-dependent moment arms, 
because we only consider small posture changes $\Delta\boldsymbol{\theta}$, we use the trapezoid integration rule such that equation~\eqref{eq:excursions} holds with $\mathbf{R} = {(\mathbf{R}(\boldsymbol{\theta}_0)+\mathbf{R}(\boldsymbol{\theta}_1))}/{2}$, where $\mathbf{R}(\boldsymbol{\theta}_0)$ and $\mathbf{R}(\boldsymbol{\theta}_1)$ are the moment arm matrices evaluated at postures $\boldsymbol{\theta}_0$ and $\boldsymbol{\theta}_1$, respectively.

We assume that each muscle acts as an ideal spring whose stiffness parameter is set by the neural command~\cite{Asatryan1965, Feldman2009, Cooke1980,Bennett1993}. We also allow each muscle to be pre-stretched from its resting length by an amount $\Delta \mathbf{l_0} = (\Delta l_1, \ldots,  \Delta l_m)^T$. The total strain energy of the system, $E$, in the presence of muscle excursions $\Delta \mathbf{s}$ is then given by
\begin{equation}\label{eq:strain energy}
\begin{split}
E & =\dfrac{1}{2} (\Delta \mathbf{s}+\Delta \mathbf{l_0})^T \mathbf{K}(\Delta \mathbf{s}+\Delta \mathbf{l_0}) ~. 
\end{split}
\end{equation}
As per the impedance, in this case stiffness, control formulation~\cite{Hogan1984}, the stiffness matrix $\mathbf{K}$ is an $m\times m$ diagonal matrix of entries $k_i$.
\begin{equation}\label{eq:Kmtx}
\mathbf{K}= 
\begin{pmatrix} 
k_{1}  & \ & 0\\  
 &  \ddots & \\
0&  &k_{m}\\ 
\end{pmatrix} ~.
\end{equation}
The corresponding forces $\mathbf{f} = (f_1, \ldots, f_m)^T$ generated by the muscles are given by
\begin{equation}\label{eq:forces}
\mathbf{f}=-\mathbf{K}(\Delta \mathbf{s}+\Delta \mathbf{l_0})~ .
\end{equation}
These muscle forces, in turn, produce torques $\boldsymbol{\tau}=(\tau_1,\ldots, \tau_n )^T$ at the joints defined by $\boldsymbol{\tau}=\mathbf{R}\mathbf{f}$.
A stable posture is achieved by satisfying the static equilibrium condition at the joints, namely, the total torque at each joint must be zero, $\boldsymbol{\tau} = \mathbf{R}\mathbf{f}=\mathbf{0}$. Slow, i.e., quasi-static, limb movements are achieved by sequentially satisfying this static equilibrium at each posture.

In the absence of external loading, if the muscles are not pre-tensioned, setting $\Delta \mathbf{l_0} = 0$ and substituting~\eqref{eq:excursions} and~\eqref{eq:forces} into $ \mathbf{R}\mathbf{f}=\mathbf{0}$ gives
\begin{equation}\label{eq:eom without pretensioning}
\mathbf{R}\mathbf{K}\mathbf{R}^T \Delta \boldsymbol{\theta} = 0 ~,
\end{equation}
which only admits the trivial solution $\Delta \boldsymbol{\theta} = 0$ for all $\mathbf{K}$ because the $n\times n$ square matrix $\mathbf{R}\mathbf{K}\mathbf{R}^T$ is full rank when $\mathbf{R}$ is full rank. Thus,
there exists only one solution at the reference posture, and other stable postures are not achievable by varying $\mathbf{K}$. In contrast, when the muscles are pre-tensioned,
one has
\begin{equation}
\label{eq:eom}
-\mathbf{R}\mathbf{K}(-\mathbf{R}^T\Delta\boldsymbol{\theta}+\Delta \mathbf{l_0}) =0 ~.
\end{equation}
 The system is thus unconditionally controllable and equilibrium postures are achievable by proper choice of muscle stiffness $\mathbf{K}$. 

The quasi-static  formulation implies that, starting from a given initial posture, the limb has to transition to ``nearby" equilibrium postures, that is to say, $\Delta \boldsymbol{\theta}$ must be small. A large value of $\Delta \boldsymbol{\theta}$ means that the limb would instantaneously jump from its current posture to a far away equilibrium posture, which violates the slow movement assumption. Therefore, in all subsequent analyses, we consider quasi-static transitions to  only nearby postures. 
In particular, we formulate: (I) an optimality problem where for a given equilibrium posture $\boldsymbol{\theta}$, we solve for optimal stiffness values that minimize the strain energy function associated with transitions to nearby postures, and (II) a reachability problem where we explore reachable equilibrium postures for given stiffness values.

\paragraph{Optimal equilibrium postures}
Starting at a posture $\boldsymbol{\theta_0}$,  a desired limb posture $\boldsymbol{\theta}$ near  $\boldsymbol{\theta_0}$ can be achieved by tuning the muscle stiffness parameters so that the vector $(-\mathbf{R}^T \Delta\boldsymbol{\theta} + \Delta \mathbf{l_0})$ in equation~\eqref{eq:eom} lies in the null space of the $n\times m$ matrix $\mathbf{R}\mathbf{K}$. 
The resulting stiffness values are not unique, in general. For each limb posture $\boldsymbol{\theta}$, equation~\eqref{eq:eom} yields a family of solutions $\mathbf{K}$, at multiple strain energy levels, for which $\boldsymbol{\theta}$ is achievable. This redundancy leads us to look for optimal stiffness values $\mathbf{K}_{\rm opt}$ that minimize the strain energy function
while satisfying the equilibrium constraint and inequality constraints on permissible muscle stiffness values, namely,
\begin{equation}\label{eq:optimization}
\begin{split}
& \mathop{\mathrm{min}}_{\mathbf{K}} \    \left[  {E}    = \dfrac{1}{2} (-\mathbf{R}^T \Delta\boldsymbol{\theta}+\Delta \mathbf{l_0})^T \mathbf{K}(-\mathbf{R}^T\Delta\boldsymbol{\theta}+\Delta \mathbf{l_0}) \right] \\[2ex]
& \mathrm{subject \  to \ }  \  \  -\mathbf{R}\mathbf{K} \left(- \mathbf{R}^T\Delta\boldsymbol{\theta}+ \Delta \mathbf{l_0}\right)=0~, \\[2ex]
& \mathrm{and}  \qquad \qquad k_{min} \le k_i \le k_{max}~. \\[1ex]
\end{split}
\end{equation}

\paragraph{Reachable equilibrium postures}
Given a specific combination of muscle stiffness parameters $\mathbf{K}$, the achievable equilibrium posture $\boldsymbol{\theta}$ can be found by solving the ``forward problem'' in~\eqref{eq:eom}, which we rewrite as $ \mathbf{R}\mathbf{K}\mathbf{R}^T\Delta\boldsymbol{\theta} = \mathbf{R}\mathbf{K}\Delta \mathbf{l_0}$.
Given that the matrix $\mathbf{R}\mathbf{K}\mathbf{R}^T$ is invertible, solutions to the forward problem are unique: for each $\mathbf{K}$, there exists one and only one $\Delta\boldsymbol{\theta}$, which in turn gives  a uniquely-defined new posture $\boldsymbol{\theta}$. 
As in the optimality problem, we only examine small changes of equilibrium posture $\Delta \boldsymbol{\theta}$.

A few remarks on the parameter values of $k_{min}$ and $k_{max}$ are in order here. $k_{min}$ is always greater than zero ensuring that the optimization problem is well-posed. Whereas leg muscles generate larger forces than finger muscles \cite{Ketchum1978, Beenakker2001, Kutch2012}, the muscle stiffness values are comparable in both, with typical values of a few hundred Newtons per meter  (N/m) \cite{Ratsep2011, Chuang2012, Bizzini2003, Mustalampi2013}. Without loss of generality, we choose a range of muscle stiffness values $k_{min} = 100$ N/m to $k_{max}=1000$ N/m, in agreement with these experimental data.
We now non-dimensionalize all parameters using a characteristic length scale $l^* = 10$ cm and a characteristic force $f^* = 100$ N.  Non-dimensional parameters enable us to obtain generic results, which we can later easily scale up or scale down to different limb sizes in both generic and subject-specific modeling.
Using the characteristic length and force scales, the dimensionless range of stiffness values is $k_{min} = 0.1$ to $k_{max}=1$. The normalized values of the pre-stretched lengths $\Delta l_i$ are all set to $0.1$ in the finger model whereas to $0.5$  in the leg model. Note that in addition to constant pre-stretch, we consider pre-stretched lengths that vary with posture, as  described in the Results section.

\section*{Results}

\begin{figure}[!t]
\begin{center}
       \includegraphics[width=0.85\textwidth]{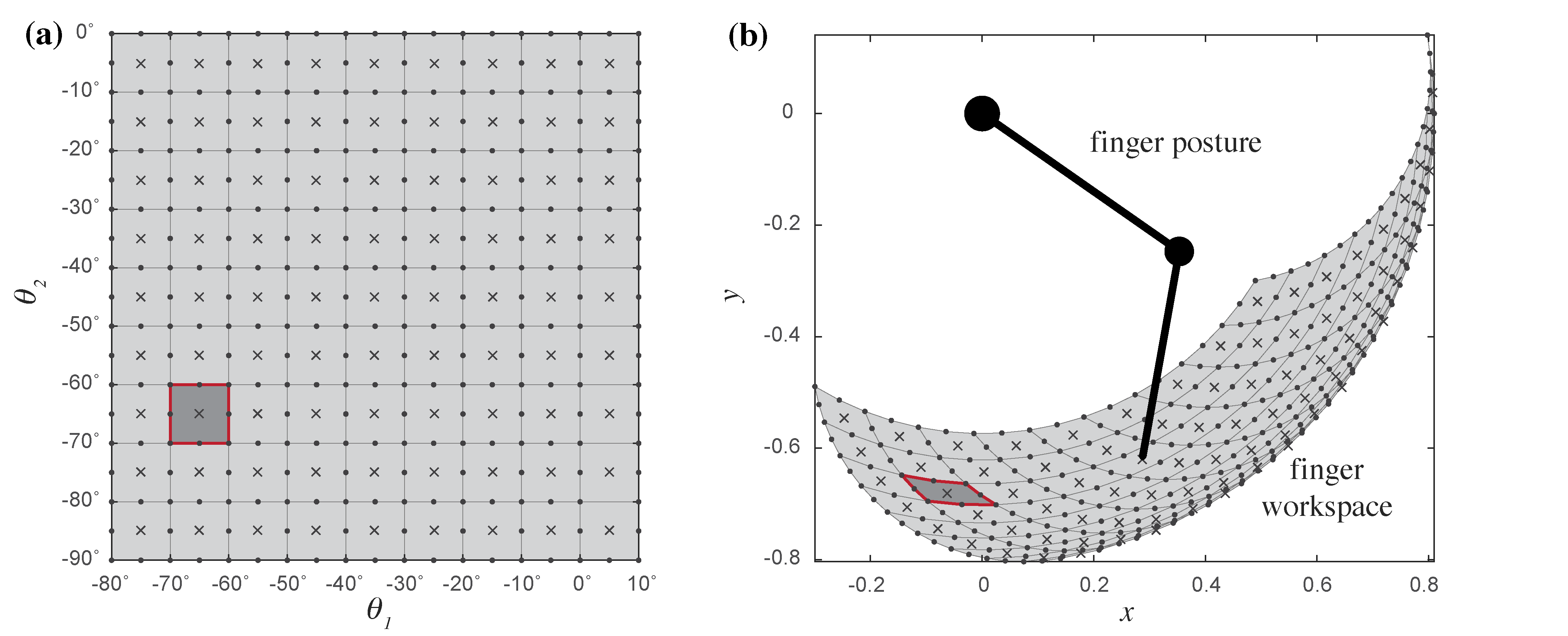}
 \end{center}
 \caption{Finger model:  (a)  configuration space and (b)  the corresponding endpoint space which deforms the squares into "diamonds."}
  \label{fig:finger vicinity locations}
\end{figure}

\paragraph{Optimal equilibrium postures} 
We consider first the simpler finger model since it is more amenable to visual representations. 
Given $\boldsymbol{\theta}=(\theta_{1},  \theta_{2})^T$, the finger's posture in the $(x,y)$-plane  follows in a straightforward way.
Its workspace $\rm W$ is the set of all points $(x_{\rm w}, y_{\rm w})$ of the $(x,y)$-plane accessible by the finger's endpoint for all 
admissible $\theta_1$ and $\theta_2$, 
\begin{equation}\label{eq:workspace}
\begin{split}
x_{\rm w} = l_1 \cos\theta_1 + l_2 \cos(\theta_2 - \theta_1) , \qquad 
y_{\rm w} = l_1 \sin\theta_1 + l_2 \sin(\theta_2 - \theta_1).
\end{split}
\end{equation}
 A depiction of the joint angles space and the workspace of the finger model is shown in Figure~\ref{fig:finger vicinity locations}.
Here, the admissible joint angles are taken to lie in the range $\theta_1 \in [-80^{\circ} ,10^{\circ}]$, $\theta_2  \in [0^{\circ}, 90^{\circ}]$. Initial postures $\boldsymbol{\theta}_0$ in this admissible range are discretized using a  $9\times 9$ regular grid, marked by $\times$. For each $\boldsymbol{\theta}_0$, we explore optimal transitions to equilibrium postures in a small square neighborhood centered at  $\boldsymbol{\theta}_0$ and of side length $10^o$. More specifically, we consider  a total of nine nearby equilibrium postures, including staying at the initial posture $\boldsymbol{\theta}_0$. We compute the optimal stiffness values for the change in joint angles $\Delta \boldsymbol{\theta}$ associated with a transition from the initial posture to each of these nine postures. We linearly interpolate these values in the neighborhood of each initial posture to construct a complete map of optimal stiffness values.

\begin{figure}[!t]
\begin{center}
       \includegraphics[width=1\textwidth]{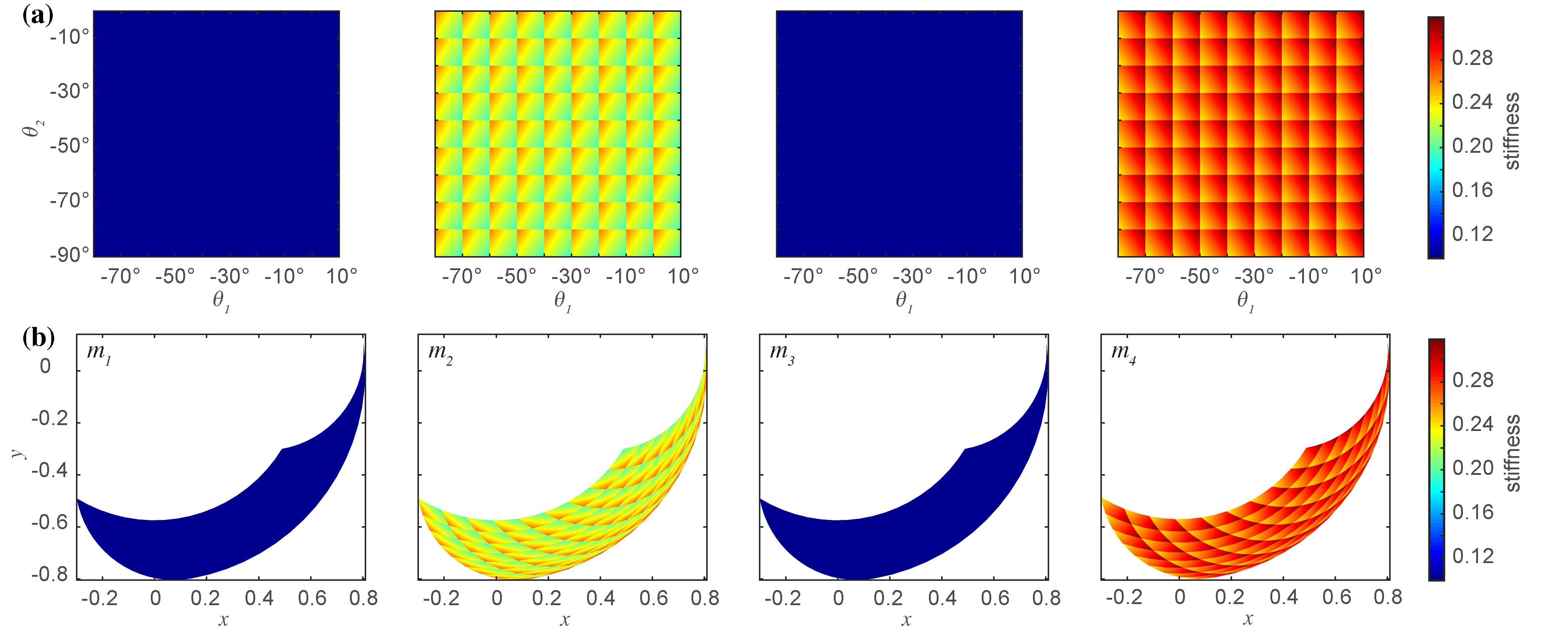}
 \end{center}
 \caption{Finger model with constant moment arms: optimal stiffness parameters needed to achieve each posture in (a)  configuration (joint angle) space, and (b)  endpoint space. The plots show the activation level of each muscle $m_i$, $i=1,\ldots, 4$, that minimize the strain energy function. Note the agonist-antagonist actions of $m_2$ and $m_4$ in the $\theta_1$-direction.} 
  \label{fig:finger stiffness vicinity constant R}
\end{figure}

Figure~\ref{fig:finger stiffness vicinity constant R} depicts the optimal stiffness values $\mathbf{K}_{\rm opt}$ for the finger model with constant moment arms. 
Muscles $m_2$ and $m_4$ exhibit higher stiffness values than $m_1$ or $m_3$, which means, considering that stiffness values are proxy to the intensity of the neural command to the muscle, that muscles $m_2$ and $m_4$ require greater neural command. A closer look at the stiffness variation within each neighborhood shows that, in the context of constant moment arms,  all neighborhoods are identical, that is to say, optimal stiffness values depend only on the change in joint angles  $\Delta \boldsymbol{\theta}$, or posture change, and not on the initial posture. Further, the stiffness variation within each neighborhood is such that $m_2$ and $m_4$ counter-balance each other in the $\theta_1$-direction -- as the stiffness of $m_2$ increases, that of $m_4$ decreases and vice-versa. This is consistent with the notion of agonist-antagonist muscle pairs.  If one muscle acts as an agonist and contracts to rotate the joint, the torque that it produces must be counterbalanced by another muscle to prevent the system from accelerating. While this language of co-contraction of agonist-antagonist pairs does not extrapolate well to multi-muscle, multi-joint systems~\cite{Valero-Cuevas2009}, the ideas that muscles produce postural equilibrium via co-contraction are well established~\cite{Hogan1984, Burdet2013}. 

\begin{figure}[!t]
\begin{center}
       \includegraphics[width=1\textwidth]{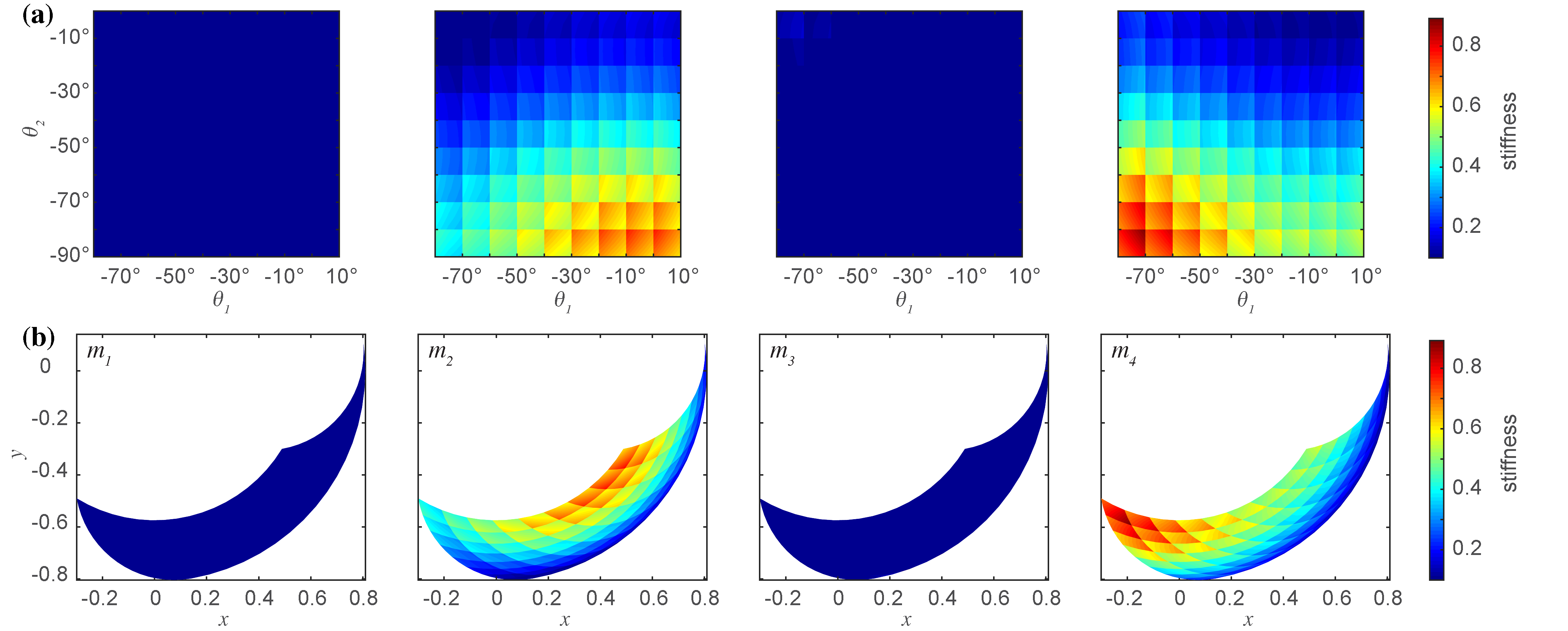}
 \end{center}
 \caption{Finger model with posture-dependent moment arms. Note the agonist-antagonist actions of $m_2$ and $m_4$ in the $\theta_1$-direction, both locally and over the whole space of joint angles.} 
  \label{fig:finger stiffness vicinity variable R}
\end{figure}

Figure~\ref{fig:finger stiffness vicinity variable R} depicts the optimal stiffness values for the case when the moment arms vary with joint angles. As in the case of constant moment arms, muscles $m_2$ and $m_4$ require higher stiffness values, and the neighborhood around each initial posture indicates that each muscle has a ``preferred" direction of higher stiffness, which is the same as in Figure~\ref{fig:finger stiffness vicinity constant R}. However, the optimal stiffness values now depend on moment arms, and thus on initial posture. Interestingly, 
the stiffness variation across the whole workspace is also characterized by the muscles $m_2$ and $m_4$ counter-balancing each other, indicating  a hierarchical structure in the way this agonist-antagonist muscle pair acts in transitions to nearby equilibrium postures and to postures across the whole workspace.

\begin{figure}[!t]
\begin{center}
       \includegraphics[width=1\textwidth]{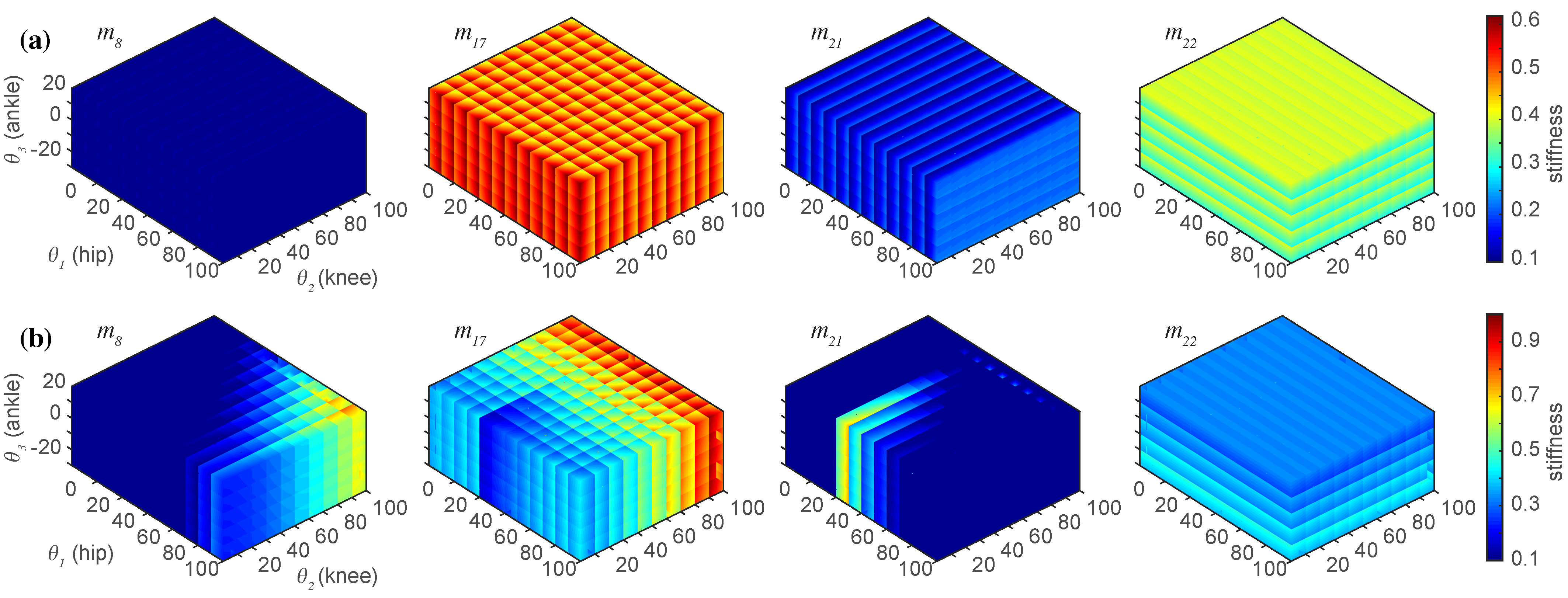}
 \end{center}
  \caption{Leg model: optimal stiffness parameters needed to achieve each posture in (a)  configuration (joint angle) space,and (b)  endpoint space. The plots show the activation level of the four highly active muscles, $m_8$ $m_{17}$, $m_{21}$ and $m_{22}$ (\textit{gluteus maximus}, \textit{rectus femoris},  \textit{tensor fasciae latae} and \textit{tibialis anterior}, respectively), that minimize the strain energy function.}  
  \label{fig:leg stiffness vicinity}
\end{figure}

In the leg model system, initial postures are taken  in the range $\theta_1 \in [-20^{\circ} ,100^{\circ}]$, $\theta_2  \in [0^{\circ}, 100^{\circ}]$ and $\theta_3  \in [-20^{\circ}, 30^{\circ}]$, and twenty-seven equilibrium postures are computed in the vicinity of each initial posture.
Figure~\ref{fig:leg stiffness vicinity}(a) shows the optimal stiffness combinations in joint angle space for constant moment arms, while Figure~\ref{fig:leg stiffness vicinity}(b) for variable, posture-dependent moment arms. For clarity, only the leg muscles that manifest the highest stiffness values are shown, namely, muscles \textit{gluteus maximus}, \textit{rectus femoris}, \textit{tensor fasciae latae} and \textit{tibialis anterior}, labeled $m_8$, $m_{17}$, $m_{21}$ and $m_{22}$, respectively. The remaining muscles, while they individually have low stiffness values,  contribute collectively to the equilibrium postures. 
As in the case of the finger model, optimal stiffness combination $\mathbf{K}_{\rm opt}$  depends on the change $\Delta \boldsymbol{\theta}$ but  is independent of the initial joint angles in the case of constant moment arms but depends on both in the case of posture-dependent moment arms.

\begin{figure}[!t]
\begin{center}
       \includegraphics[width=0.8\textwidth]{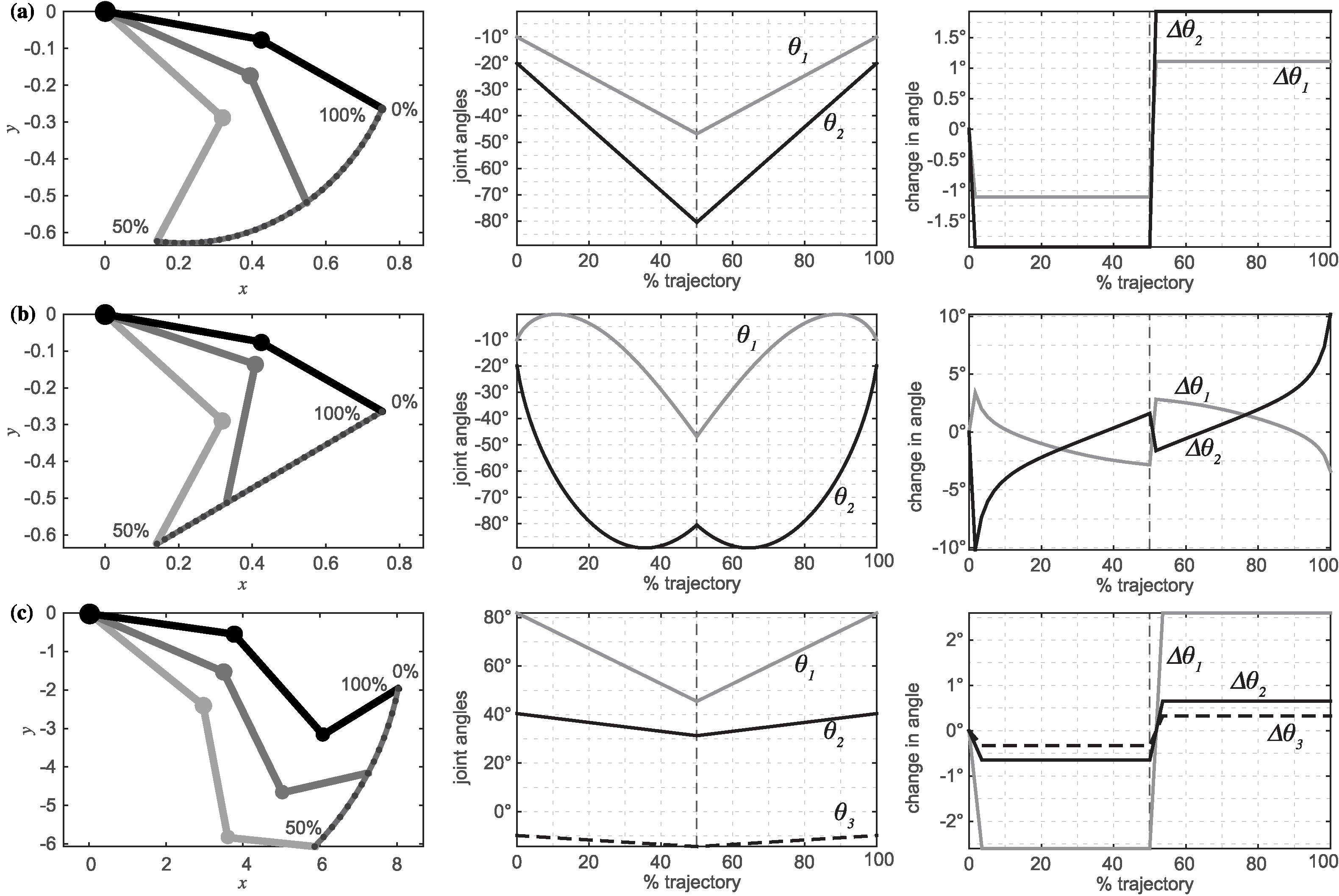}
 \end{center}
 \caption{Endpoint trajectories for (a) and (b) finger model and (c) leg model.}
  \label{fig:trajectories}
\end{figure}

\paragraph{Quasi-static trajectories} We consider quasi-static trajectories where the limb endpoint is required to slowly trace, back and forth, curved and straight lines in the workspace as depicted in Figure~\ref{fig:trajectories}. We solve for the optimal muscle stiffness combinations required by the limb to perform this back and forth motion. To this end, the quasi-static trajectories are discretized, and the optimization problem in equation~\eqref{eq:optimization} is solved sequentially along the discrete trajectory.

\begin{figure}[!t]
\begin{center}
       \includegraphics[width=0.8\textwidth]{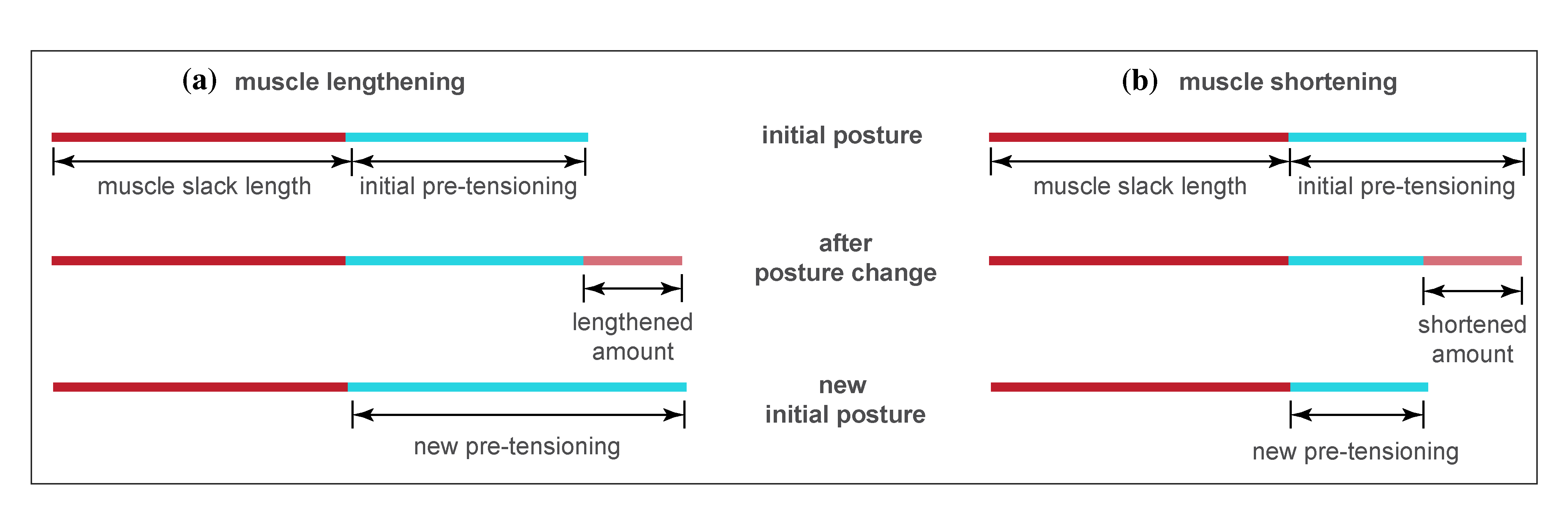}
 \end{center}
 \caption{Variable pre-tensioning update rule: the pre-tensioning is history-dependent, taking into account the muscle lengthening (a) or shortening (b) that took place in the previous posture change.}
  \label{fig:variablepretensioning}
\end{figure}

We investigate the effect of constant versus variable muscle pre-tensioning.  
Constant pre-tensioning means that the system is ``memoryless"  because muscle pre-tensioning does not depend on the previous muscle configuration. Variable pre-tensioning is defined according to the update rule depicted in Figure~\ref{fig:variablepretensioning}, where the pre-tensioning value at each posture is equal to that at the previous posture plus the change in muscle length that occurred as a result of moving from the previous to the current posture; thus the pre-tensioning has ``memory." 

\begin{figure}[!t]
\begin{center}
       \includegraphics[width=1\textwidth]{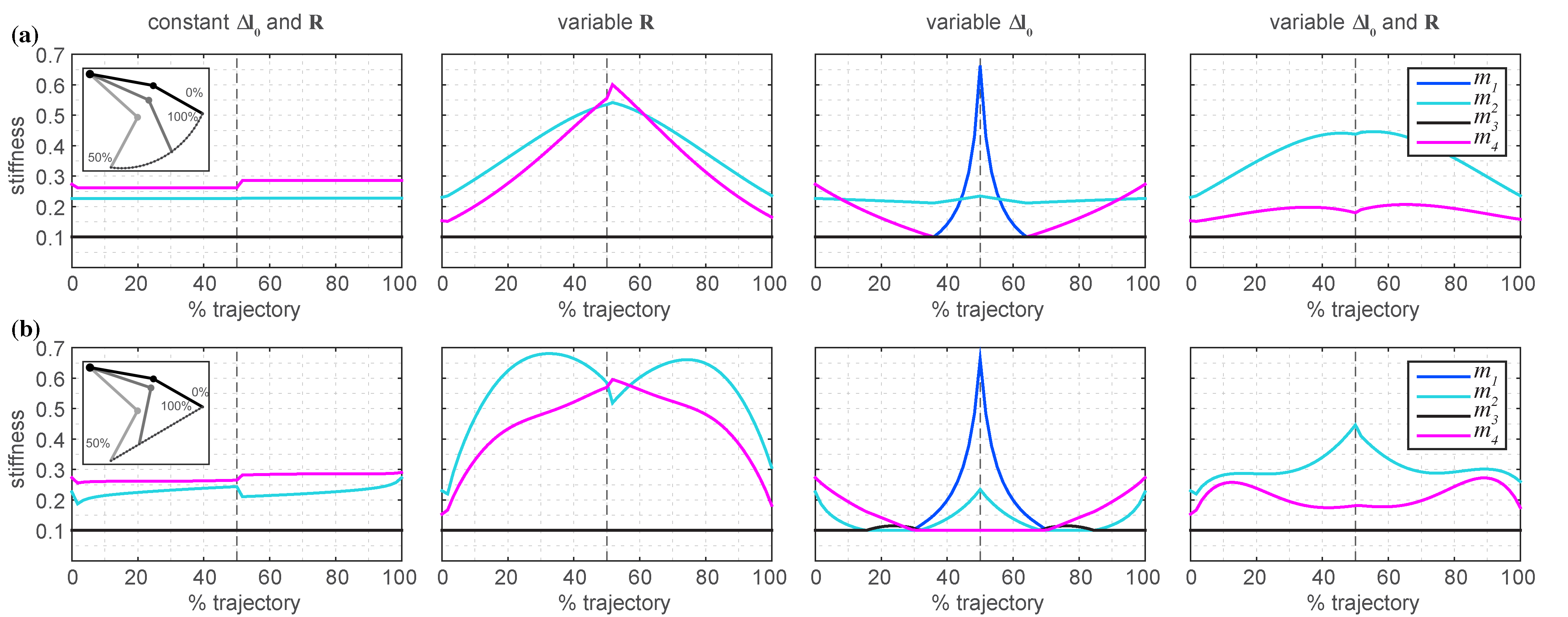}
 \end{center}
 \caption{Finger model: optimal stiffness values for the quasi-static trajectories shown in Figure~\ref{fig:trajectories}(a,b).} 
  \label{fig:finger stiffness trajectories}
\end{figure}

\begin{figure}[!t]
\begin{center}
       \includegraphics[width=1\textwidth]{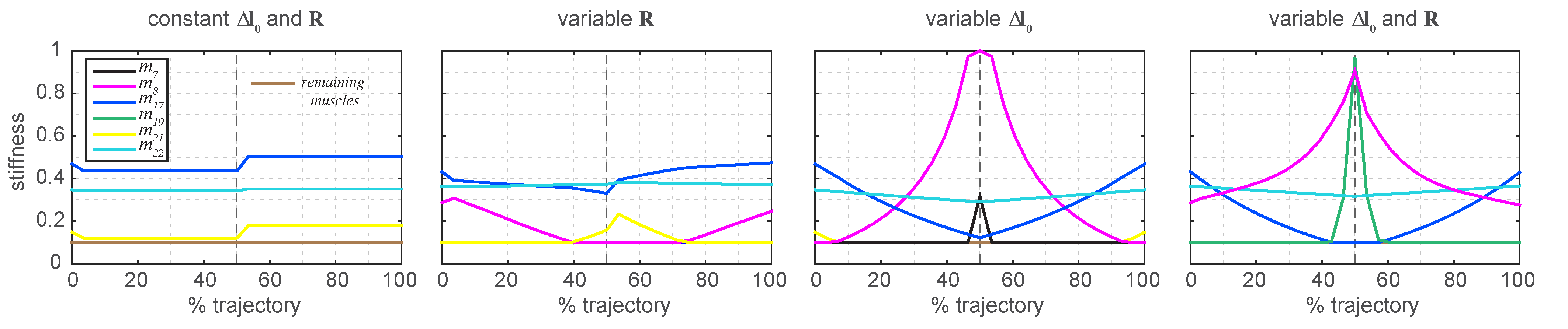}
 \end{center}
 \caption{Leg model: optimal stiffness values for quasi-static curved trajectory shown in Figure~\ref{fig:trajectories}(c). The most active muscles during this trajectory are the \textit{gluteus maximus} ($m_7$ and $m_8$), \textit{rectus femoris} ($m_{17}$), \textit{biceps femoris long head} ($m_{19}$), \textit{tensor fasciae late} ($m_{21}$) and \textit{tibialis anterior} ($m_{22}$).}
  \label{fig:leg stiffness trajectories}
\end{figure}

Figure~\ref{fig:finger stiffness trajectories} depicts the optimal stiffness values for the finger model tracing the trajectories of Figure~\ref{fig:trajectories}(a,b) for all combinations of moment arms and pre-tensioning scenarios, whereas  Figure~\ref{fig:leg stiffness trajectories} shows similar results for the leg model tracing the curve of Figure~\ref{fig:trajectories}(c). 
In both models, the optimal stiffness values are characterized by a local jump as the limb reverses its motion to trace the trajectory backward. The jump in the stiffness of certain muscles gets attenuated in the case of variable moment arms, but all jumps get smoothed out completely when the muscle pre-tensioning is changing continuously with posture. In these cases,  one observes the relaxation of the antagonist muscles during the activity of the agonist, and  the opposite effect is clearly seen during the return portion of the movement. For instance, the \textit{gluteus maximus} ($m_8$) and \textit{rectus femoris} ($m_{17}$) in the leg model act against each other to complete the trajectory.

\paragraph{Reachable equilibrium postures} We explore the reachability or feasibility of tracing certain trajectories within the limb's workspace given the inherent constraints on the physiological precision in the neural control of muscle stiffness. The precision with which muscle activation can be controlled is not infinite; it is limited by the mechanisms of recruitment and rate coding used to control muscles. Thus, the space of possible stiffness values $\mathbf{K}$ is not infinitely smooth. Here, we systematically sample the stiffness space at different precision levels, i.e., degrees of discretization granularity $\Delta k$. For each precision level $\Delta k$, we calculate the set of admissible departures ${\Delta \boldsymbol{\theta}}$ from a reference posture, and consequently the set of reachable (feasible) postures $\boldsymbol{\theta}$.  Our goal is not to minimize the strain energy at each reachable posture, but to find the set of nearby reachable postures as a function of the precision with which individual muscle stiffnesses can be controlled. Minimizing the strain energy at each reachable posture introduces additional constraints, making it even more challenging to reach a desired posture at a given stiffness precision level.

We discretize the range of stiffness values from $k_{min}$ to $k_{max}$ using constant increments $\Delta k$, producing $M=(k_{max}-k_{min})/\Delta k + 1$ discrete stiffness values for each muscle. For  $m$ muscles, this discretization generates $M^m$ different stiffness matrices $\mathbf{K}$, which amounts essentially to a uniform sampling of the $m$-dimensional muscle stiffness space. For example, setting $\Delta k = 0.05$ in the finger model amounts to $19^4$ distinct $\mathbf{K}$ matrices. For each $\mathbf{K}$, we solve equation~\eqref{eq:eom} to compute $\Delta\boldsymbol{\theta}$ starting from a reference posture $\boldsymbol{\theta_0}$, and use~\eqref{eq:workspace} to compute the reachable endpoint location $(x_{\rm w},y_{\rm w})$. We constrain the full set of $M^m$ reachable points to those that are in the vicinity of the reference posture.

\begin{figure}[!t]
   \centering
        \includegraphics[width=1\textwidth]{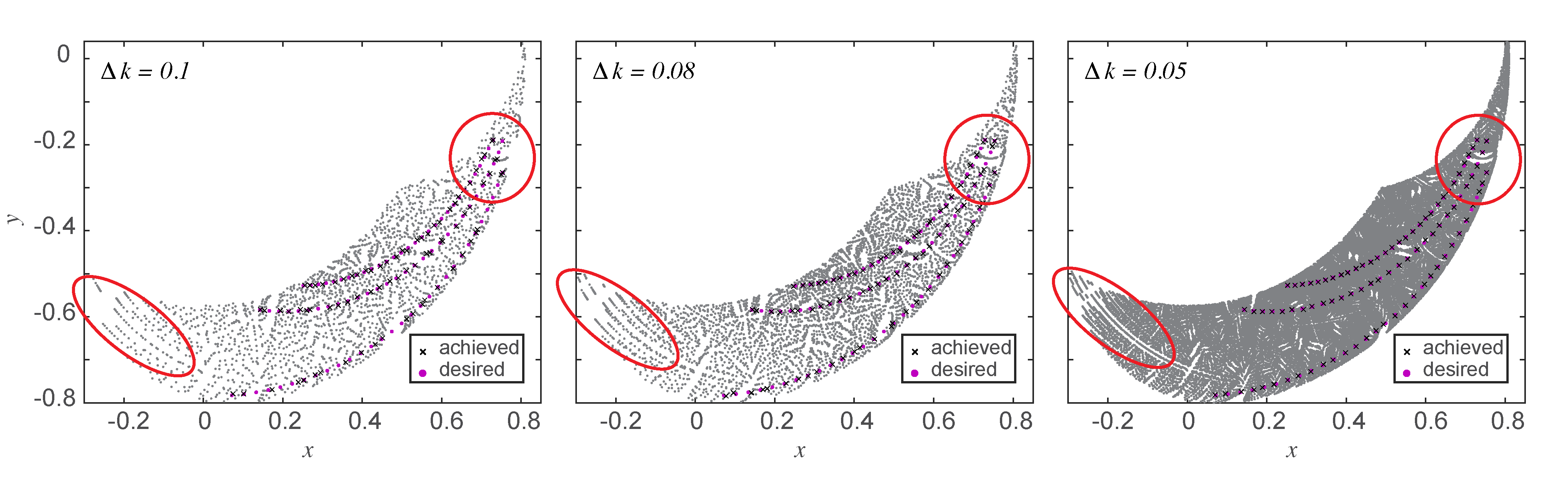}
         \caption{Finger model: achievable versus desired trajectories overlaid on top of the reachable postures. The level of stiffness precision affects the reachability. As precision increases (from left to right), the endpoint trajectories become smoother and closer to the desired trajectories. Discontinuities in the set of reachable points vary nonlinearly in the workspace, e.g., regions highlighted in red.} 
               \label{fig:finger reachability trajectories}
\end{figure}

For illustration purposes, we investigate the ability of the finger model to trace multiple trajectories for a given $\Delta k$.
Starting from the initial posture, we sequentially transition to the next point along the trajectory by locating the nearest admissible reachable posture. 
Figure~\ref{fig:finger reachability trajectories} shows the desired versus reachable endpoint trajectories for three precision levels $\Delta k=0.1, 0.08, 0.05$, superimposed on the set of all reachable point in the limb workspace. The reachable paths show discontinuities. These discontinuities decrease as the resolution of muscle stiffness increases (going from $\Delta k = 0.1$ and $\Delta k = 0.05$). The discontinuous behavior arises from unreachable desired postures, in which case the limb settles for the nearest admissible posture from the set of reachable postures. Further, we observe that discontinuities are spatially localized, implying that the demands on the precision in muscle stiffness varies nonlinearly in the limb workspace.

\section*{Discussion} 

We developed a novel mathematical formulation for slow and accurate limb movements using the framework of \textit{driftless} mechanical systems, and considered the neural control of muscles to change the muscles' stiffness parameters. We used this framework to address three fundamental questions in slow limb movements: (1) can slow limb movements be achieved by modulating the muscle stiffness parameters only? (2) what is the role of muscle tone? and (3) how are these movements affected by limits on the precision with which muscle stiffness can be controlled? We demonstrated that muscle stiffness control is sufficient to produce accurate and slow limb movements, but only when the muscles are pre-stretched, as in the case of natural muscle tone. We probed the limb's ability to trace prescribed trajectories under constraints on muscle stiffness precision. We identified discontinuities in trajectory tracking. By comparing different resolutions of muscle stiffness, we found that high resolution of muscle stiffness is key for the limb to improve its accuracy in tracking a trajectory. 

 The discontinuous nature of slow finger movements was reported in numerous experimental studies \cite{Vallbo1993, Darling1994}. Several studies attributed these discontinuities to neural 
sources while others held responsible peripheral stretch reflexes \cite{Gross2002, Williams2009, Evans2003}. Recent reports seem to suggest an inter-related network that 
includes both the brain and the periphery \cite{Williams2009, Williams2010}. Our work supports the latter view of a combined neuro-mechanical origin of the discontinuities and presents an alternative and novel explanation of the mechanism mediating these discontinuities, namely, the precision in muscle stiffness. 
Muscle stiffness is known to have two components: (i) short-range or inherent stiffness of the muscle coming from sarcomere mechanics, motor unit recruitment and rate coding, and tissue properties; and (ii) reflex-driven stiffness coming from the stretch-reflex response mediated by spinal and transcortical afferent circuits~\cite{Cui2008, Hu2011, Burdet2013}. Pathological disruption of either mechanism could degrade the level and precision of muscle tone and muscle stiffness. 
One example is abnormally high muscle tone, or hypertonia, which is defined as abnormally increased resistance to externally imposed movement about a joint, and may be caused by spasticity, dystonia, rigidity, etc.~\cite{Sanger2003}. Other examples include pathologic tremor and bradykinesia (slow movements) in people with Parkinson's disease~\cite{Kopin1993, Jankovic2008, Bettray2013}, and stroke~\cite{Krebs1999}. 
Our results suggest that degradation in the precision of muscle stiffness would exacerbate the inherent non-smoothness of slow movements in tendon-driven limbs and provide alternative and complementary peripheral explanations for the aforementioned clinical symptoms.

We made a number of simplifying assumptions. The assumption that small finger muscles and larger lower limb muscles have similar stiffness range is based on recent reports that relatively small finger muscles, such as the \textit{extensor digitorum} or \textit{flexor carpi radialis}, have stiffness values~\cite{Chuang2012} similar to a larger leg muscle, such as the \textit{rectus femoris} (around $300 N/m$ at rest)~\cite{Mustalampi2013}. We neglected gravity in both the finger and the leg model but we obtained similar results in the presence of gravitational effects (not shown for brevity). 
We considered the initial pre-tensioning length $\Delta \mathbf{l_0}$ as a constant or  posture-dependent parameter, and assumed that only the muscle stiffness is under direct neural control. In future extensions of this work, we will explore the case where the pre-tensioning values are also regulated by neural drive and consequently change the muscle force. 
\\

\section*{Acknowledgments}
We thank Prof. Jason Kutch for sharing the leg muscle model. 
SB and FVC were partially supported by the grants NIH R01AR050520, NIH R01AR052345, NIDRR H133E080024 and NSF EFRI-COPN 0836042, and SB and EK by the grants NSF CMMI-0644925 and NSF CCF-0811480. The content is solely the responsibility of the authors and does not necessarily represent the official views of the National Institutes of Health.

\bibliography{R_Strain_Energy_References}
\end{document}